# Cryogen-free variable-temperature Kelvin probe force microscopy for probing local chemical potential in a graphene heterostructure

Namkyung Lee[1]\*, Seungwon Jung[1]\*, Baeksan Jang[1], Sangwook Ha[1] and Joonho Jang[1]

Department of Physics and Astronomy, and Institute of Applied Physics, Seoul National University, Seoul 08826, Korea

\* These authors equally contributed to this work

We report the development of a variable-temperature Kelvin probe force microscopy (KPFM) system capable of stable and highly sensitive operation over a wide temperature range based on a GM-cooler-based cryogen-free cryostat. The system incorporates a custom-designed phase-locked loop and automatic gain control, along with passive vibration isolation, enabling precise measurements of local chemical potential even under cryogenic conditions. We demonstrate the performance of this setup by measuring hBN encapsulated monolayer graphene (MLG), revealing spatially resolved electronic inhomogeneities and charge puddles. Our measurements clearly capture temperature-dependent variations in the chemical potential near the charge neutrality point (CNP), consistent with the linear band dispersion of MLG and interaction-driven renormalization of Fermi velocity. This work highlights the robust sensitivity and stability of our system, making it a versatile local probe of quantum phases in van der Waals heterostructures.

## I. INTRODUCTION

Recent studies on electronic phases in low-dimensional nanostructures have revealed a variety of intriguing physical phenomena. In particular, two-dimensional systems such as graphene and transition metal dichalcogenides (TMDs), which exhibit competing topological and correlated electronic orders at low temperature, offering new platforms to discover exotic quantum phases.[1–3] However, most conventional studies rely on bulk transport measurements, which only provide spatially averaged information. Such methods make it difficult to directly probe essential physical properties, such as the local distribution of electronic phases and topological electronic phenomena near the sample edges.[4,5] To overcome these limitations, local measurement techniques based on scanning probe microscopy (SPM) have gained increasing attention as complementary electrical characterization techniques

SPM enables fine measurement of local electronic properties using a probe as a sensitive sensing element. Owing to the probe's small size and non-invasive nature, SPM has the advantage of directly and quantitatively measuring physical phenomena at the nano-scale—that complements conventional bulk transport measurements.[6–8] Among SPM techniques, there are a few examples which are sensitive to electronic signals, including scanning single-electron transistor (sSET)[9,10], scanning tunnelling microscopy (STM)[11] and Kelvin probe force microscopy (KPFM).[12–14] Each technique offers powerful advantages for a specific type of measurement, but also has clear limitations. For example, the sSET provides high sensitivity capable of detecting a single electron, but can operate only at cryogenic temperatures. In the case of STM, it provides excellent spatial resolution and direct access to local density of state (LDOS), but requires extremely demanding vibration isolation and ultra-high vacuum conditions. In contrast, KPFM is based on a highly sensitive mechanical oscillator with a relatively robust mechanism of maintaining the tip-sample distance, and therefore enables stable measurement under various surface conditions. Crucially, it not only serves as a window to access the work function of material systems but also offers important insight into the local chemical potential of electronic phases within a material system by spatially mapping it across a sample.

Therefore, variable-temperature KPFM with high sensitivity would provide a powerful tool for studying delicate quantum phenomena in 2D material systems. Especially due to the remote sensing capability of KPFM, even electrical signals from 2D systems encapsulated by hBN dielectric layers can be measured, ensuring a pristine local environment for 2D systems. [15–17] However, reliable operation in various temperatures with high sensitivity has always been challenging for SPM. In a KPFM system with a cryogen-free cooler, in particular, it is crucial to isolate a high-Q cantilever from vibrations of the cryocooler and to implement a robust feedback mechanism to maintain the cantilever tip in proximity to the sample surface.

In this paper, we introduce a home-built KPFM system that operates reliably over a wide temperature range using a GM-cooler-based cryogen-free cryostat. The system is designed to suppress mechanical vibration across a broad frequency spectrum, enabling stable measurements at ambient and cryogenic temperatures. We demonstrate the versatility and sensitivity of our setup by performing illustrative measurements on MLG, the prototypical 2D system, and present precise measurements of local chemical potential distribution near the CNP and high-resolution visualization of the charge puddle profile. Our results demonstrate a scanning microscopy setup with the potential to study various other 2D materials with marked improvements over previous approaches, and achieve high electrical potential sensitivity and high spatial resolution.

## II. RESULT AND DISCUSSION

### A. *Design and operation of the variable temperature KPFM*

KPFM is a SPM technique capable of measuring the local chemical potential with nanometer scale spatial resolution. It is based on the principle that, when two conductive materials with different work functions come into electrical contact, electrons move between them to equalize their electrochemical potential, developing an electrostatic potential difference—called contact potential difference $U_{CPD}$ - between the two materials. When DC bias $V_{tip}$ is applied to the tip and AC bias $V_{AC}sin(\omega_e t)$ is applied to the sample as shown in **Fig. 1** and **Fig 3(b)**, electrostatic potential difference $U = (U_{CPD} - V_{tip}) +$

$V_{AC}sin(\omega_e t)$ is developed at the tip-sample capacitance $C_{eff}$, which is kept constant along with the tip-sample distance z of a tens-of-nanometer scale by atomic force microscopy (AFM) feedback loop.

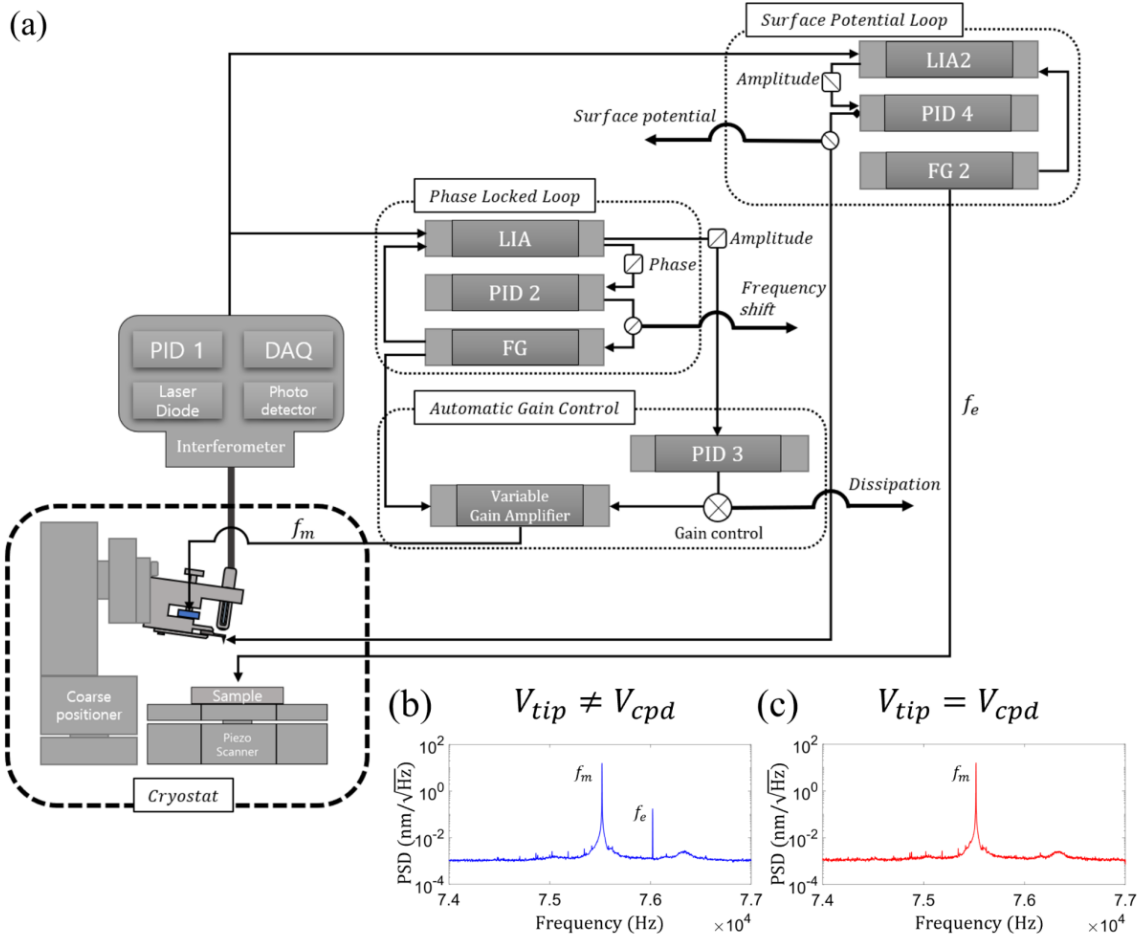

**Fig 1. (a) An overview of the custom-built KPFM system.** (a) Surface topography is measured by frequency modulation with the resonance frequency of the cantilever $f_m$, and $U_{CPD}$ is measured by amplitude modulation with an excitation frequency $f_e$. $f_e$ is set ~ 500Hz off-resonance from $f_m$. A PLL with AGC tracks $f_m$ in real time, keeping the tip stably close to the sample surface even under vibration. (b,c) PSD of the cantilever's displacement. The resulting peak is visible in (b) nullified by Kelvin feedback Loop (PID4) in (c). The cantilever has a Q factor of ~15000.

The resulting electric force $F = -\frac{1}{2}(\partial C_{eff}/\partial z)U^2$ mechanically drives the cantilever. since first harmonic ($\omega_e$) part of the force $F$ is linearly proportional to $U_{CPD} - V_{tip}$, KPFM find the $V_{tip}$ value that nullifies this first harmonic force, which equals $U_{CPD}$. Since $U_{CPD}$ equals the chemical potential difference between the tip and the sample, KPFM enables to retrieve the local chemical potential profile across the sample surface, also providing insights into the electronic state of the bulk material.

For the AFM feedback loop, the displacement of the cantilever is monitored by a fiber-based interferometer with a collimation manifold, in a fully vacuum and cryogenic-temperature compatible design (see **Fig. 1**). The relative position of the sample stage and the cantilever-collimator manifold is controlled by xyz-axis coarse positioners (SmarAct cryogenic positioning systems) and a piezo scanner (Attocube ANSxyz100). The three-axis coarse positioners and a fine scanner integrated into our system provide a travel range of 5 mm and a maximum scan range of 35 um, respectively, even under cryogenic conditions. This enables effective sample positioning and scanning. The automatic gain control (AGC) unit maintains a constant amplitude of the cantilever's motion and enhances the stability of the frequency shift–based feedback loop by controlling the excitation voltages fed into a piezo actuator[18].This AGC consists of an analogue multiplier (AD633) and an analog PID controller built with op-amps, while a similar PID controller was also built for the main loop filter of the phase-locked loop (PLL).

At the core of the signal processing workflow in our design, is a self-designed PLL. This PLL system is integrated with a lock-in amplifier (LIA), a voltage controlled amplifier (VCA), and an analog PID controller (PID2). The phase detector (LIA), we employed a LIA (SRS SR865A) with a frequency coverage up to 4 Mhz, which enabled precise and accurate phase extraction at the cantilever's resonance frequency $f_m \sim 75\ kHz$. The LIA was configured with 3rd-order low-pass filter with time constant of 30 us to ensure the bandwidth sufficient to follow the change of the $f_m$. This configuration allowed us to effectively extract $f_m$, while suppressing the second and higher harmonics. For the VCA, we utilized the frequency modulation functionality of a commercial function generator (Siglent SDG2082X) by applying the PID controller's output as an external modulation signal, enabling fast and stable frequency tuning. The analog PID (PID2) was built using OP27 operational amplifiers, selected for their 8 Mhz bandwidth, 2.8V/us slew rate and exceptionally low noise floor of 3 nV/√Hz, and has proportional and integral components and an adder circuit, allowing real-time correction of the phase error and precise tuning of PID gains. As a result, it tracks the resonance of the cantilever in real-time with 3kHz bandwidth. The loop filter of PLL is a combination of analogue and digital filters, which effectively suppress high-frequency noise and achieve a frequency resolution of approximately 5 mHz with ±300 Hz tunable ranges. Taking spatial images of the $U_{CPD}$ values, we employ a one-pass method, where the surface topography is measured using the frequency-modulation method at $f_m$, while the CPD is acquired by continuously nulling the amplitude-modulation signal at $f_e$ slightly detuned (~500 Hz) from $f_m$.

### B. *Characterization of vibration isolation and detection sensitivity*

Accurate characterization of the system noise is a key metric in evaluating the overall reliability of the SPM system. KPFM utilizes mechanical detection of forces based on a high-Q mechanical oscillator and thus is prone to vibrational noises from external sources. In particular, in an environment of low temperature and high vacuum, the quality factor of the cantilever increases significantly, making it more susceptible to vibrations and environmental disturbances. Our KPFM is designed to be accommodated in a cryocooler-based cryogen-free cryostat, which generates mechanical vibrations from circulating compressed gas, therefore we had to implement a proper vibration isolation and noise suppression strategy.

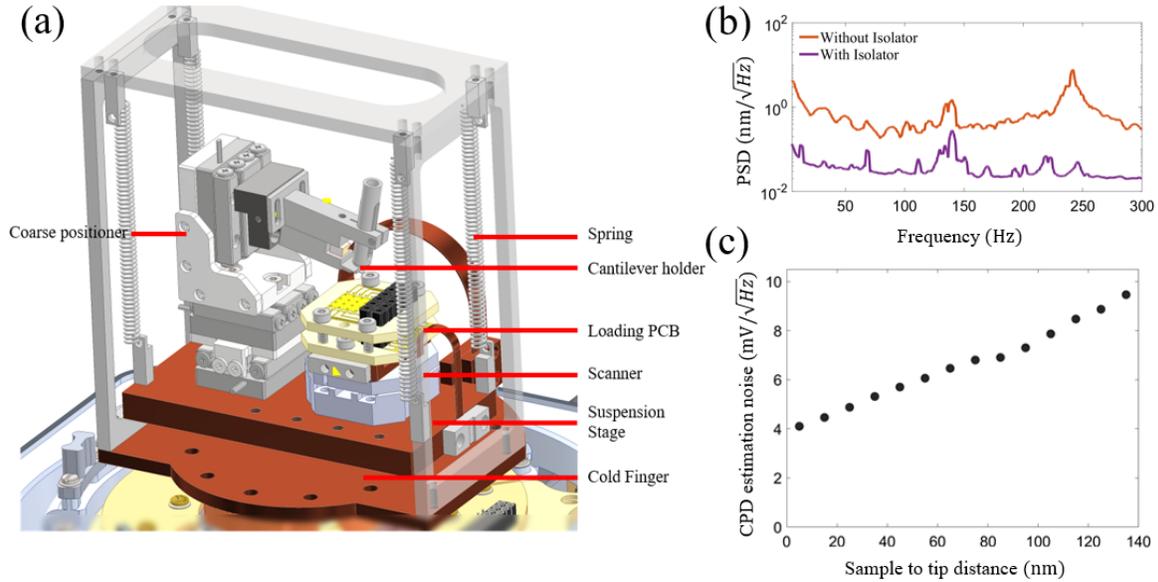

**Fig 2. Characterization of vibration isolation and sensitivity of KPFM** (a) A CAD rendering of the KPFM system. An additional passive vibration isolator–composed of a soft mechanical spring ($k$ =16 mN/m) and a soft copper braid-based heat sink–is implemented to effectively reject the external vibration. (b) Power spectral density (PSD) of tip sample vibration measured using a Fabry-Perot interferometer under conditions : near the cryocooler without any isolation (orange) with an additional passive vibration isolator installed in the chamber (purple). The vibration amplitude is significantly reduced from 20.79 $nm_{RMS}$ to 0.89 $nm_{RMS}$ after implementing our vibration isolation system. (c) Estimated CPD noise calculated using error propagation method from the slope of the V-shaped curve in the EFM amplitude. The noise show a linear dependence on the tip-sample distance, and reaches approximately 4.8 $mV/\sqrt{Hz}$ at 5 nm over 30 nm thickness hBN distance under 0.1Vpp AC bias.

    While some degree of vibration isolation is provided by the cryostat (Montana instruments S200) itself, to further suppress the vibrations, we designed and set up an additional passive vibration isolation stage as shown in **Fig. 2(a)**, which effectively suppressed GM-cooler induced vibration in a wide frequency range, reducing the RMS noise amplitude. To analyze the system's vibrational characteristics, we monitored in real time the distance between the sample stage and the cantilever-holding manifold using a high-bandwidth optical interferometer. The data is expected to represent the effective noise strength in a vibrational mode of the setup that directly affects the distance between the tip and a sample - the most critical interface for KPFM operation. As shown in **Fig. 2(b)** the results of measurements were plotted with

power spectral density (PSD) over a frequency range of 3 Hz to 300 Hz, before and after installation of the vibration isolation stage. The comparison reveals the vibration amplitude is reduced from 20.79 nm$_{RMS}$ to 0.89 nm$_{RMS}$. This substantial improvement enables stable KPFM measurements in a cryogenic condition under the influence of the cryocooler's strong vibration.

Another important component for evaluating the performance of a KPFM system is how accurately it can measure $U_{CPD}$. The measurement noise of $U_{CPD}$ directly affects the accuracy of determining the CNP, which in turn has a significant impact on analysing the electronic properties of 2D materials such as MLG. Therefore, a quantitative evaluation of the $U_{CPD}$ noise level is essential. As aforementioned, the first harmonic part of the electric force arising from the capacitive coupling between the tip and the sample is linearly proportional to $U_{CPD} - V_{tip}$, making it directly sensitive to $U_{CPD}$. When this first-harmonic signal (often referred to as the EFM amplitude) is plotted as a function of $V_{tip}$, it yields a line whose slope serves as a sensitive indicator of the CPD noise level. During KPFM measurements, the tip-sample DC bias is continuously adjusted to make the cantilever amplitude at the excitation $f_e$ monitored via a second LIA (LIA2) to be zero using a feedback loop (PID4); the tip-sample voltage is recorded as CPD (See also PSD plots in **Fig, 1**). Using this method, we can show that our system achieves a CPD sensitivity of approximately $4.8 \ mV/\sqrt{Hz}$ at a tip-sample separation of 5 nm across a 30 nm-thick hBN, under 0.1 Vpp AC bias, which places it as one of the most sensitive cryogenic KPFM measurements in the literature.

### C. *Measurement of a hBN-encapsulated monolayer graphene device*

Further demonstrating the performance of our KPFM design, for detecting local potential landscape and serving as a probe of local electronic properties with high spatial resolution, we performed benchmark measurements on a material system with unique DOS, MLG. It was previously reported that scanning SET (sSET) is a capable tool to visualize spatial non-uniformity of the CNP of MLG, resulting from the presence of local charge puddles[19,20]. To quantitatively evaluate the sensitivity and spatial resolution of our design of KPFM in comparison to sSET, we prepared a MLG device fabricated with hBN dielectric encapsulation as shown in **Fig. 3(a,b)**. The device consists of a MLG sandwiched by top and bottom hexagonal boron nitride (hBN) layers, The top and bottom layers are 27 nm and 38 nm thick, respectively, as measured by AFM. This heterostructure is placed on a graphite flake, which serves as the bottom gate, and is subsequently stacked onto a 300 nm-thick SiO$_2$ layer atop a p-doped Si substrate. Pre-patterned Ti/Au electrodes provide ohmic electrical contacts to the MLG.

A major strength of KPFM lies in its capability to determine the CNP is MLG by analysing the $U_{CPD}$ as a function of the $V_{bg}$. Due to its sensitivity to the LDOS, KPFM can be used to pinpoint the local CNP position $(V_{bg}, U_{CPD})=(V_0, U_0)$. where the $U_{CPD}$ exhibits the maximum slope. We demonstrate that the $U_0$ can be spatially resolved, while also evidencing the stable operation and high sensitivity of our KPFM design through the acquired 2D scanning images. In **Fig. 3(c)** we scanned a region of 3.75 $\mu m$ by 3.75 $\mu m$ at $T = 30 \ K$ and found that the topography of the hBN surface is not uniform. **Fig. 3(d)** presents the $U_{CPD}$ distribution measured at 0 V. This demonstrates that our KPFM is capable of detecting potential variations with sensitivity on the order a few $mV/\sqrt{Hz}$. In **Fig.3 (e)**, the distinct spatial variation of $U_0$ is clearly resolved, thanks to the robust operation and high detection sensitivity of our setup.

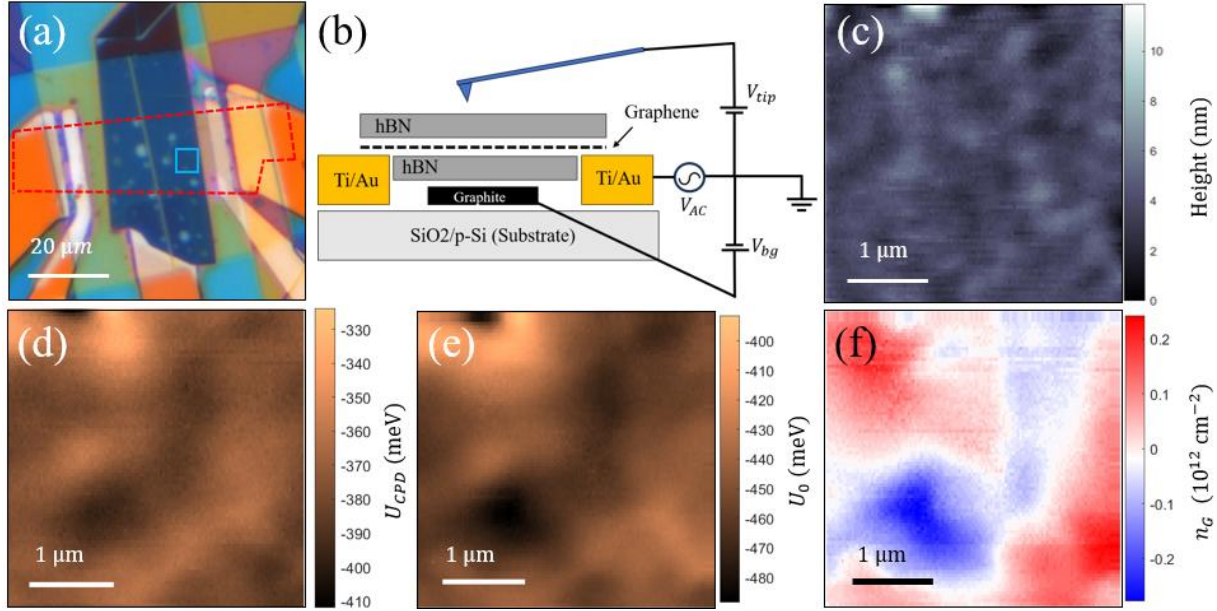

**Fig. 3. Spatial mapping of the chemical potential and carrier density in the MLG device.** (a,b) Optical image and schematic of the MLG device are shown respectively. During KPFM measurement of the region marked in (a) as a blue square with the backgate voltage of $V_{bg} = 0\,V$, spatial distribution of (c) topography and (d) $U_{CPD}$ are retrieved simultaneously. Each position has a difference $V_{bg}$ offset $V_0$ and $U_{CPD}$ offset $U_0$ for its CNP. (e) and (f) show the spatial profile of $U_0$ and retrieved carrier density offset CNP ~ $(C_g/e)(V_0 - [V]_{avg})$, respectively.

This variation reflects the particular DOS near the Dirac point of the MLG, even though it is located underneath a 27 nm-thick hBN dielectric. This variation reflects the particular DOS near the Dirac point of the MLG, even though it is located underneath a 27 nm-thick top hBN dielectric. While the AFM topography in **Fig. 3(c)** achieves a spatial resolution of about 30 nm-limited mainly by the tip-sample separation-the stray electric field between the sample and the tip that broadens the range of electrical interaction and presence of the top hBN dielectric layer limit the KPFM spatial resolution to roughly 100nm, as shown in **Fig. 3(d, e)**.

Moreover, the approximate carrier density $n_G$ of MLG at the global CNP ~ $(C_g/e)(V_0 - [V]_{avg})$ shows a spatially varying pattern of electron-hole puddles as shown in **Fig. 3(f).** We attribute the irregular surface topography and the related chemical potential suppression to the existence of residues at one of the interfaces of the heterostructure and charge accumulation that occurred during the sample fabrication process (see also **Supplementary section S2**). This moderate spatial variation of the carrier density on the scale of several $10^{11}\ cm^{-2}$ is comparable to that of the reference[21].

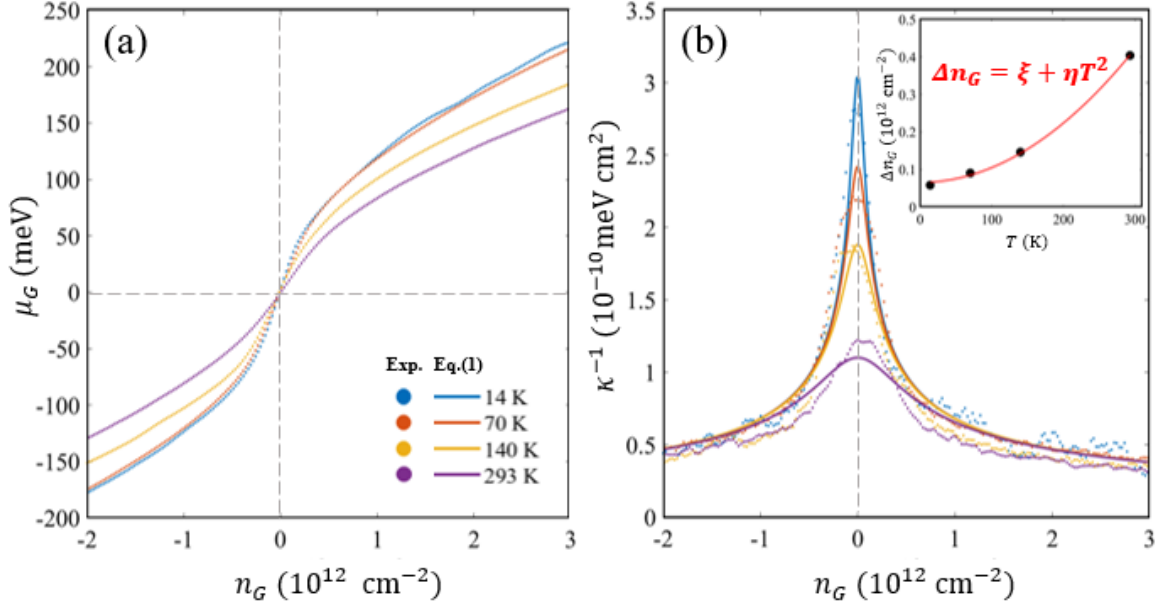

**Fig 4. Local chemical potential measurements of a hBN encapsulated MLG device** (a) Local chemical potential $\mu_G$ as a function of carrier density $n_G$, measured at varying temperatures $T$. (b) Inversion compressibility $\kappa^{-1} = \frac{\partial \mu_G}{\partial n_G}$ as a function of $n_G$, showing a thermal broadening near the CNP. Experimental data (marked with dots) were fit with theoretical Eq. 1 (marked with solid lines), and extracted $\Delta n_G$ was fit as a function of temperature $T$ as $\Delta n_G = \xi + \eta T^2$.

The spatially-scanned images of the values thus serve as sensitive experimental indicators for local electrical environments across the sample, and showcase that our KPFM instrumentation is capable of performing high-resolution profiling of local electronic properties of hBN-encapsulated 2D materials devices with high stability and accuracy.

To extract more quantitative insight from the spatially varying $U_{CPD}$ data, we analyze how the local $U_{CPD}$ can be systematically converted into local chemical potential and carrier density. (see also Supplementary Materials Fig. S2) Converting raw data of KPFM - local CPDs ($U_{CPD}$) and back-gate voltages ($V_{bg}$) - to the information of the local chemical potential ($\mu_G$) as a function of the charge carrier density ($n_G$) of the MLG requires a careful analysis. If we define the chemical potential of the MLG to be zero at CNP, it is evaluated as $\mu_G/e = U_{CPD} - U_0$. Because the CPD measured with KPFM in a material is equal to the local chemical potential plus a constant material-dependent CPD, we attribute $U_0$ (the potential value for CNP) to represent the local charge distribution due to residues such as interfacial hydrocarbons that are not accountable by the charge density of the MLG. Then, the carrier density of the MLG is given by $n_G = (C_g/e)(V_{bg} - V_0 - \mu_G/e)$, where $C_g$ is the geometric capacitance per unit area between the back gate and the MLG. The relationship between the chemical potential and the carrier density are expected to provide essential information on the band dispersion and LDOS of the system under investigation, MLG in this case.

To further analyse the extracted information and test our variable-temperature local chemical potential detection scheme, we measured $U_{CPD}$ as a function of $V_{bg}$ of 15 local spots in the same 2um by 2um area of the device, converted axes to $U_{CPD}$ and $V_{bg}$ to $\mu_G$ and $n_G$, and averaged to obtain the data at 4 different temperatures in **Fig. 4(a)**. Then, the inverse compressibility $\kappa^{-1} \equiv (\partial \mu_G / \partial n_G)$ is numerically evaluated as a function of $n$ and plotted in **Fig. 4(b)**. Taking the derivative of each chemical potential curve reveals a pronounced peak, whose position and shape exhibit a clear temperature dependence, with which we extract local band dispersion information. While the Fermi velocity of a MLG is known to be approximately $0.95 \times 10^6 \ m/s$[21], in a simple model, the electron-electron interaction actually modifies the electronic dispersion near CNP[22,23]; the value of the Fermi velocity increases logarithmically in the low carrier density limit and at low temperatures. This modified dispersion changes the density dependence of the chemical potential and we are able to fit $\kappa^{-1}$ curves by the following equation [24]

$$\kappa^{-1} = \frac{d\mu_G}{dn_G} = \frac{\sqrt{\pi}\, \hbar v_F}{2(n_G^2 + (\Delta n_G)^2)^{1/4}} \left(1 + \frac{\alpha c}{8\epsilon_m v_F} \ln\left(\frac{n_D}{\sqrt{n_G^2 + (\Delta n_G)^2}}\right)\right) \qquad (1)$$

where $\alpha$ denotes a fine-structure constant, and $\hbar$ is reduced planck constant, $\epsilon_m = 8$ denotes the effective dielectric constant of the medium in which the MLG is embedded where electron-electron interactions inside MLG are screened by hBN substrate and self-screening in MLG[23], $\Delta n_G = \xi + \eta T^2$ accounts for fluctuation of $n_G$ due to electron-hole puddles ($\xi$) and thermal excitation ($\eta$), and $n_D = 6 \times 10^{14} \ cm^{-2}$ denotes a density corresponding to the energy cut off 3 eV[21]. The extracted parameter values $\xi = 6.55 \times 10^{10} \ cm^{-2}$ and $\eta = 3.95 \times 10^6 \ cm^{-2} K^{-2}$ are comparable with those of the reference[24]. A comparison between room-temperature and cryogenic measurements reveals that thermal excitation near the Dirac point is significantly suppressed at low temperatures, thereby allowing the effects of the renormalized band dispersion to be clearly observed and quantitatively extracted. Overall, these results demonstrate that our variable-temperature KPFM is a highly effective tool for quantitatively probing temperature-dependent changes in the electronic structure of MLG.

### III. CONCLUSIONS

In this study, we introduce a KPFM system capable of operating in a cryocooler-based cryogen-free cryostat. To ensure high sensitivity even at cryogenic temperatures, the system integrates custom-designed PLL/AGC feed-back circuits and a passive mechanical vibration isolation structure, enabling precise measurement of the local chemical potential. Using this setup, we observed distinctive variations in the chemical potential near the CNP in MLG. By analysing the temperature-dependent slope changes of the CPD curves, we quantitatively evaluated the interaction-induced renormalization effect on MLG's unique dispersion and DOS. Our design marks an advancement of the KPFM instrumentation towards a versatile tool to investigate spatially-resolved, temperature-dependent information of electronic phases in 2D materials heterostructures with high measurement sensitivity.

## ACKNOWLEDGEMENT

This work was supported by the National Research Foundation of Korea grants funded by the Ministry of Science and ICT (Grant Nos. 2019R1C1C1006520, RS-2020-NR049536, and RS-2023-00258359), the Institute for Basic Science of Korea (Grant No. IBS-R009- D1), SNU Core Center for Physical Property Measurements at Extreme Physical Conditions (Grant No. 2021R1A6C101B418), and Creative-Pioneering Researcher Program and Samsung DS Basic Research Program (Project No. 0409-20230298) through Seoul National University.

## CONFLICT OF INTEREST

Authors declare no conflict of interest.

## DATA AVAILABILITY

Data presented and used to support the results and analysis in this work can be provided upon a reasonable request to the corresponding author.

**Supplementary Materials**

Additional experimental details and theoretical analysis are provided in the supplementary information. This includes the sample fabrication procedure using a dry transfer method for hBN-encapsulated monolayer graphene (Fig. S1), an effective circuit model for interpreting KPFM measurements in terms of local chemical potential and carrier density (Fig. S2), and the detailed calibration method used to extract physical quantities such as chemical potential ($\mu$) and carrier density (n) from raw CPD vs. backgate voltage data (Fig. S3). These sections also clarify how geometric and quantum capacitance contribute to the overall system response and discuss the origin of spatial inhomogeneity in the extra CPD signals. Figure S4 shows typical images acquired in the process of locating samples. Figure S5 displays an example image of protrusions on the surface of samples possibly originating from bubbles or residues.

**SUPPLEMENTARY MATERIALS**

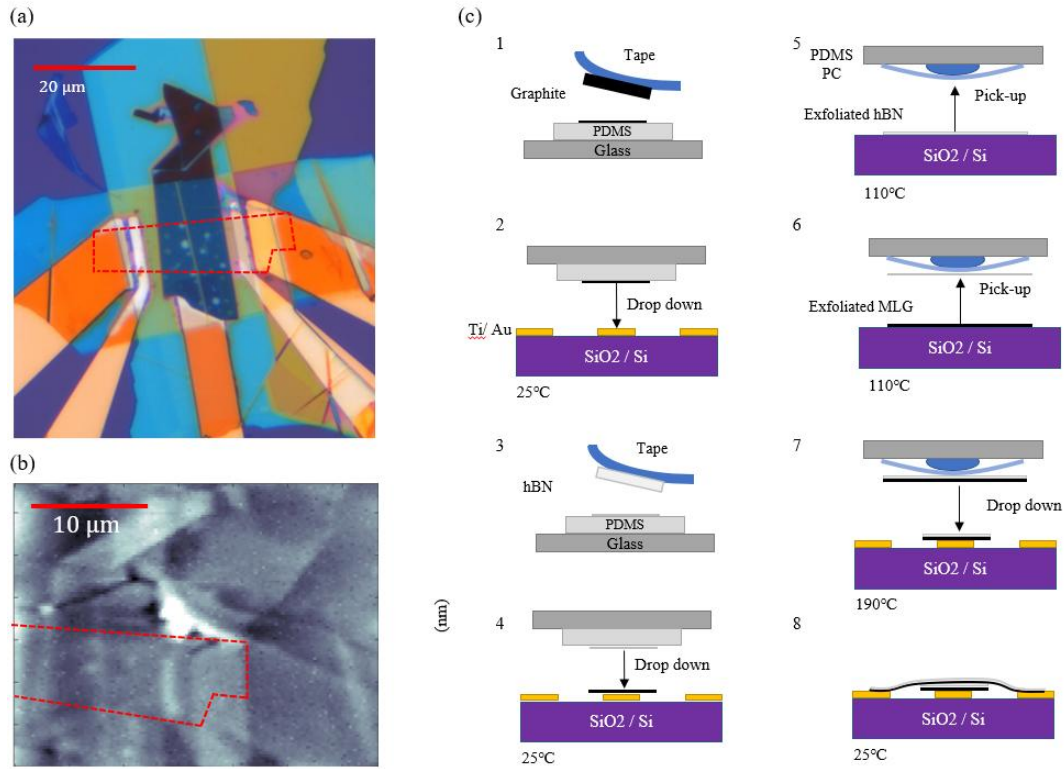

**Fig S1. (a) Optical micrograph of our sample (b) Fabrication process of our MLG heterostructure (c) Topographic image of the area measured under 14 K**

*S1. Sample fabrication*

For high quality sample fabrication, we employed a method of hBN capsulation with dry transfer. First, the graphite was exfoliated and transferred onto Si/SiO2 (300 nm) substrate pre-patterned with Ti/Au (5nm/50nm) electrodes using photolithography – using a GelPak PDMS stamp, and used as a local backgate. Next, exfoliated hBN (38 nm) was transferred in the same method to work as a bottom insulating layer, providing electrical isolation from the graphite backgate. Separately exfoliated MLG was then picked up together with the hBN (27 nm) as top insulating layer using a PC/PDMS-based dry stamp, and precisely aligned and dropped down on the pre-patterned electrode to form the final heterostructure. The completed sample was mounted onto a custom-designed cryogenic PCB to ensure stable operation under vacuum and low-temperature conditions, and was integrated into the KPFM systems.

*S2. Accurate determination of DOS (chemical potential vs. carrier density) near Dirac point*

In 2D materials devices, the carrier density is tuned capacitively by the back-gate voltage. Then, the value of the capacitance used to charge carriers is the $C_{eff}$ formed between the back-gate and the sample, which can be modeled as a total capacitance of two capacitances connected in series as $C_{eff} = (C_g^{-1} + C_Q^{-1})^{-1}$. Here, $C_g$ is the geometric capacitance, which can be estimated as $(30/t)$ fF/$\mu m^2$ where $t$ nm is the thickness of hBN separating the graphite back-gate and the sample, and $C_Q$ is the quantum capacitance, which is governed by the DOS within the electronic system of the MLG (or any system under investigation).

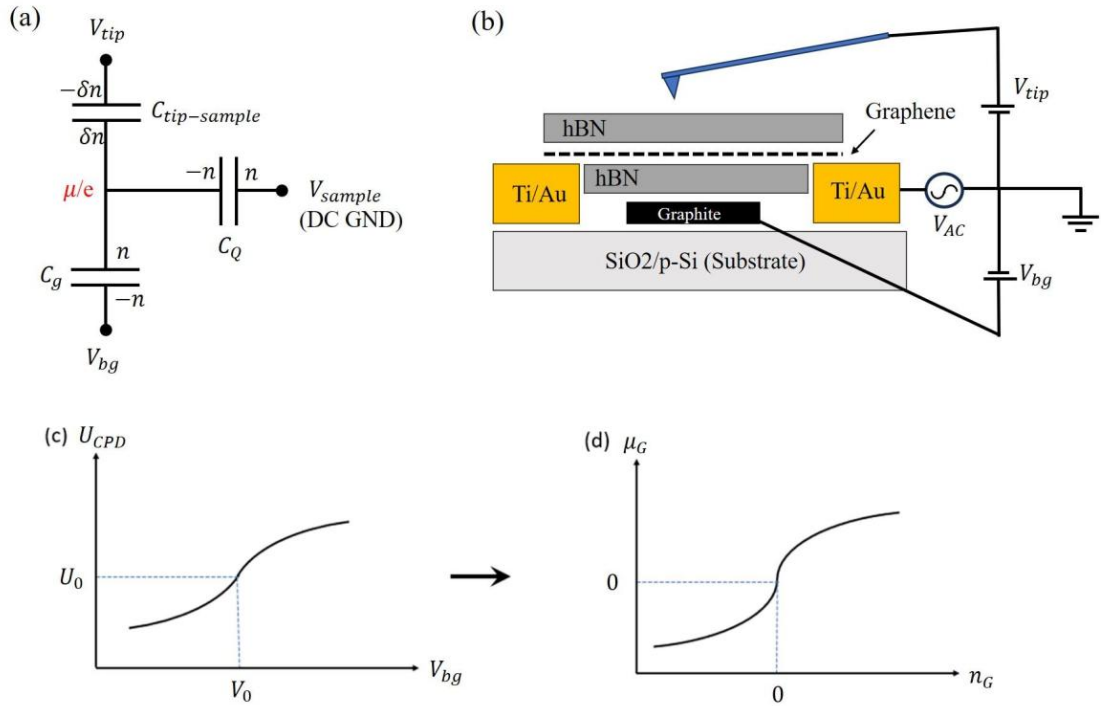

**Fig S2. Effective circuit diagram for the analysis of KPFM's detection of the chemical potential $\mu$ of MLG heterostructures**

In **Fig. S2(b)**, we present an effective model to describe the KPFM operation. The electrostatic force arising from the charge distribution within the sample can be described by a capacitive force by the electrostatic potential difference between the tip and the sample, induced by the chemical potential difference between them. Thus, by nulling the capacitive force for the tip-sample capacitance, which makes $\delta n$ zero, the measured CPD becomes equal to the chemical potential of the sample. Then, the charge density is simply given by the geometric capacitance multiplied by the difference between the back-gate voltage and the chemical potential of the sample. The density $n_G = (C_g/e)(V_{bg} - V_0 - \mu_G/e)$ and the chemical potential $\mu_G/e = U_{CPD}$ are determined by the two voltages $V_{bg}$ and $U_{CPD}$ in the experiments.

For extracting the dispersion information from a 2D material, however, we need to redefine the chemical potential $\mu_G$ and the carrier density $n_G$ referenced to the minimum of the conduction band, or the CNP in the case of MLG, and we assign $\Delta V_{bg} = V_{bg} - V_0$ and $\Delta U_{CPD} = U_{CPD} - U_0$, where $V_0$ and $U_0$ are the corresponding voltages for CNP (see **Fig. S2(c)**), and then $\mu_G/e = \Delta U$ and $n_G = (C_g/e)(\Delta V_{bg} - \Delta U_{CPD})$ (see also **Fig. S2(d)**). For example in the case of a MLG, without velocity renormalization, the relationship between the chemical potential and the density is expected to be given by $\mu_G = (\hbar v_F/e^2)\sqrt{\pi n_G}$.

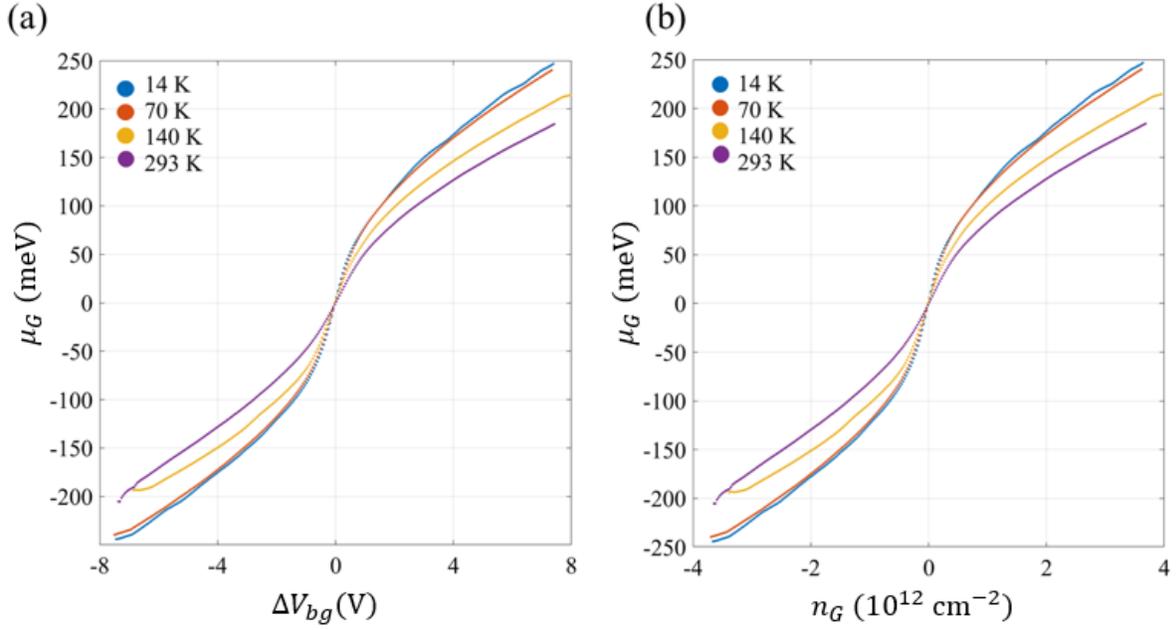

**Fig S3.** (a) $\mu_G$ vs $\Delta V_{bg}$ of KPFM measurement. (b) is converted data via $n_G = (C_g/e)(\Delta V_{bg} - \Delta U_{CPD})$

Meanwhile, the values $V_0$ and $U_0$ are quantities that can provide the information of charge and energetic inhomogeneity and thus local electrical environment of the sample in principle. The shifts in $V_0$ can be understood as charge density modulation of the 2D system due to an electrostatic potential influence from remote external sources, while the variations of $U_0$ likely indicates the contact potential difference coming from foreign materials in contact with the 2D system. In our current experiment, we believe they are dominated by hydrocarbon or water residues introduced in the fabrication steps.

In **Fig. S3**, the raw data measured in $U_{CPD}$ and $V_{bg}$ and the calibrated data in terms of the chemical potential and the charge density are compared. Small but important modifications of the data due to the calibration process are visible near CNP at low temperatures.

*S3. Additional scanning images*

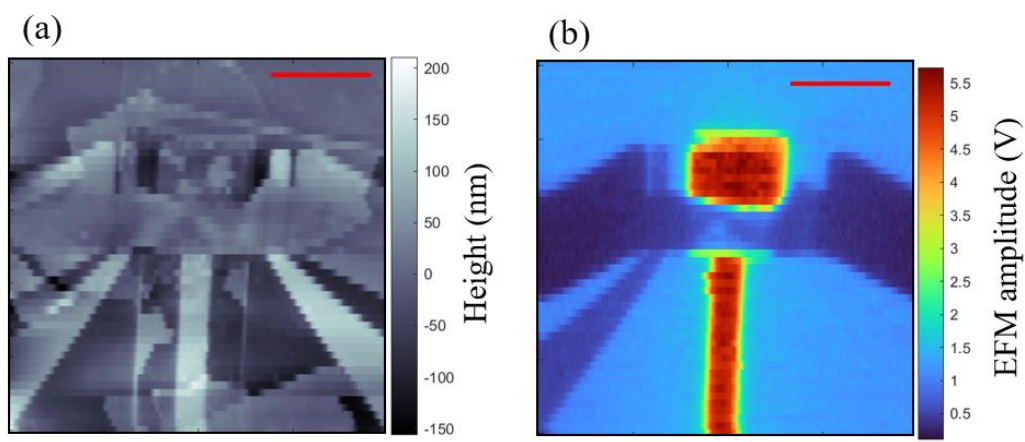

**Fig S4. Typical images to identify a sample's location** (a) AFM topography (b) EFM measurement. The scale bar is 10 $\mu m$. These are images of another sample different from the one in the main text, but show typical images acquired in the process of locating samples.

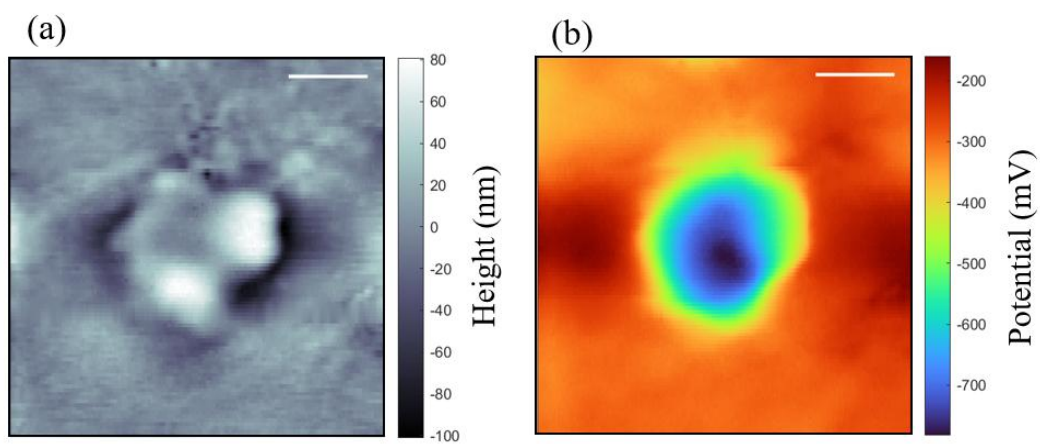

**Fig S5. A measurement of a large object above the MLG** (a) AFM topography (b) KPFM measurement. The scale bar is 1 $\mu m$. These display example images of protrusions on the surface of samples possibly originating from bubbles or residues.